\documentstyle[12pt]{article}

\begin{document}

\thispagestyle{empty}

\hfill \parbox{45mm}{{UTF-367/96} \par Jan 1996} 

\vspace*{15mm}

\begin{center}
{\LARGE Updating the Theoretical Analysis of the}

\smallskip

{\LARGE Weak Gravitational Shielding Experiment.}
        
\vspace{22mm}

{\large Giovanni Modanese}%
\footnote{e-mail: modanese@science.unitn.it}

\medskip

{\em I.N.F.N. -- Gruppo Collegato di Trento \par
Dipartimento di Fisica dell'Universita' \par
I-38050 POVO (TN) - Italy}

\bigskip \bigskip

\medskip

\end{center}

\vspace*{10mm}

\begin{abstract}
The most recent data about the weak gravitational shielding 
produced recently through a levitating and rotating HTC 
superconducting disk show a very weak dependence of the 
shielding value ($\sim 1 \%$) on the height above the disk. We 
show that whilst this behaviour is incompatible with an intuitive 
vectorial picture of the shielding, it is consistently explained 
by our theoretical model. The expulsive force observed at the 
border of the shielded zone is due to energy conservation.

\medskip
\noindent
74.72.-h High-$T_c$ cuprates.         

\noindent
04.60.-m Quantum gravity.

\bigskip 

\end{abstract}
 
The measurements of Podkletnov et al.\ of a possible weak gravitational 
shielding effect \cite{p1,p2} have been repeated several times and under 
different conditions by that group, with good reproducibility,
including results in the vacuum. In the forthcoming months
other groups will hopefully be able to confirm the effect independently.
While the Tampere group was mainly concerned with  
obtaining larger values for the shielding, studying its dependence 
on numerous experimental parameters and testing new materials
for the disk, in the future measurements it will be important
to obtain more exact data, including detailed spatial field maps. 
The theoretical model suggested by us \cite{m1} is still evolving,
although at a fundamental level; a more detailed account
appears elsewhere \cite{m2}.

Let us recall in short the main features of the experiment. A HTC
superconducting disk or toroid with diameter between 15 and 30 $cm$,
made of $YBa_2 Cu_3 O_{7-x}$, is refrigerated by liquid helium in a
stainless steel cryostat at a temperature below $70 \ K$. The
microscopic structure of the material, which plays an important
role in determining the levitation properties and the amount
of the effect, is described in details in the cited works.

The disk levitates above an electromagnet and rotates by
the action of lateral alternating e.m. fields. Samples of different
weight and composition are placed over the disk, at a distance
which can vary from a few $cm$ to 1 $m$ or more (see below). A
weight reduction of about 0.05\% is observed when the disk is
levitating but not rotating; the weight loss reaches values about
0.5\% when the disk rotates at a frequency of ca.\ 5000 $rpm$. 
If at this point the rotating fields are switched off, the
sample weight remains decreased till the rotation frequency
of the disk decreases. On the other hand, if the rotation
frequency is decreased from 5000 to 3500 $rpm$ using the solenoids
as breaking tools, the shielding effect reaches maximum values
from 1.9 to 2.1\%, depending on the position of the sample
with respect to the outer edge of the disk.

This effect, if confirmed, would represent a very new and 
spectacular phenomenon in gravity; namely, as well known, there
has never been observed any conventional gravitational shielding 
up till now, up to an accuracy of one part in $10^{10}$, and General 
Relativity and perturbative Quantum Gravity exclude any measurable 
shielding \cite{m1}. Our temptative theoretical explanation is based
on some properties of non-perturbative quantum gravity. We have
shown that the density field $|\phi_0|^2$ of the Cooper pairs
inside the superconductor or, more likely, the squared gradient
$(\partial_\mu \phi_0)^* (\partial^\mu \phi_0)$ may act as localized 
positive contributions to the small negative effective gravitational 
cosmological constant $\Lambda$; if the sum turns out to be positive 
in a certain four-dimensional region, a local gravitational singularity 
arises there, affecting the gravitational propagators and thus the 
interaction potential (between the Earth and the samples, in
this case).

To sketch our model -- although not rigorously -- we could say 
that there is an "anomalous coupling" between the mentioned 
density $|\phi_0|^2$ or the squared gradient $(\partial_\mu \phi_0)^* 
(\partial^\mu \phi_0)$ and the gravitational field, and
that the net result is to partly "absorb" the field. We expect
that only in some regions of the superconductor the density
$|\phi_0|^2$ or the squared gradient will be strong enough and
that the inhomogeneities of the material and the pinning centers
will be crucial in determining such regions. Since the gravitational 
field is attractive, its "absorption" requires energy from the 
outside. This means that there must be some mechanism external to
the disk which {\it forces} $|\phi_0|^2$ or $(\partial_\mu \phi_0)^* 
(\partial^\mu \phi_0)$ to take high values. This is caused in the 
experiment by the action of the external electromagnetic field and 
by the disk rotation.

The dependence of the shielding effect on the height, at which the
samples are placed above the superconducting disk, has been recently 
measured up to a height of ca.\ 3 $m$ \cite{p3}. No difference in the 
shielding value has been noted, with a precision of one part in
$10^3$. It is also remarkable that during the measurement at 3 $m$ height
the sample was placed in the room which lies above the main laboratory,
on the next floor; in this way the effect of air flows
on the measurements was greatly reduced. For the used 500 $g$ sample 
the weight loss was ca. 2.5 $g$.

Such an extremely weak height dependence of the shielding is in 
sharp contrast with the intuitive picture, according to which the
gravitational field of the Earth is the vectorial sum of the
fields produced by each single "portion" of Earth. In the absence
of any shielding, the sum results in a field which is equivalent 
to the field of a pointlike mass placed in the center of the Earth;
this can be checked elementarily by direct integration or invoking
Gauss' theorem and the spherical symmetry. But if we admit that
the superconducting disk produces a weak shielding, the part of
the Earth which is shielded lies behind the projection of the
disk as seen from the sample, i.e., within an angle $\theta$ about
the vertical direction, such that $\ \tan \theta = h$, where $h$ is
the sample height over the disk measured in units of the 
disk radius. (For simplicity we suppose now the sample to be 
centered above the disk.)

In order to obtain the shielding effect as a function of $h$, 
taking into account this geometrical factor, one must integrate 
the Newtonian contribution $\cos \phi /R^2$ over the intersection 
between the Earth and the cone defined by $\phi<\theta$.
We have done this for the values $h=1,2,3,4,6,8,10$, through a 
Montecarlo algorithm. We took into account the higher density 
of the Earth's core ($\rho_{core} \sim 2 \rho_{mantle}$; $r_{core}
\sim (1/2) r_{Earth}$; it is straightforward to insert more accurate
values, but the final results change very little); we also computed 
analytically the contribution of the tip of the cone, from the 
Earth's surface to the Earth's core, in order to reduce the 
fluctuations in the Montecarlo samplings for small $R$.
\footnote{For the detailed algorithm and figures please ask
the author at the e-mail address above.}
The resulting values were the following:

\begin{verbatim}

  h     shielding/maximum-shielding  
  =================================
  1           0.62  +/- 0.02
  2           0.34  +/- 0.01
  3           0.18  +/- 0.01
  4           0.102 +/- 0.003
  6           0.050 +/- 0.002
  8           0.029 +/- 0.001
  10          0.018 +/- 0.001

\end{verbatim}

This strong height dependence is clearly incompatible with the mentioned 
experimental data, which instead seem to indicate that in the 
shielding process all the mass of the Earth behaves effectively 
as if it would be concentrated in one point.

In our theoretical model this property arises in a natural way.
We employ a quantum formula which expresses the static gravitational
interaction energy of two masses $m_1$ and $m_2$ in terms of an
invariant vacuum expectation value, namely \cite{m3}
\begin{eqnarray}
  E & = & \lim_{T \to \infty} - \frac{\hbar}{T}
  \log \frac{\int d[g] \, \exp \left\{ - \hbar^{-1} \left[
  S[g] + \sum_{i=1,2} m_i \int_{-\frac{T}{2}}^{\frac{T}{2}} dt \,
  \sqrt{g_{\mu \nu}[x_i(t)] \dot{x}_i^\mu(t) \dot{x}_i^\nu(t)}
  \right] \right\}}{\int d[g] \, \exp \left\{ - \hbar^{-1} S[g] \right\} }
  \label{ciao} \\
  & \equiv & \lim_{T \to \infty} - \frac{\hbar}{T} \log 
  \left< \exp \left\{ - \hbar^{-1} \sum_{i=1,2} m_i 
  \int_{-\frac{T}{2}}^{\frac{T}{2}} ds_i \right\} \right>_S
\label{bella}
\end{eqnarray}
where $g$ has Euclidean signature and $S$ is the gravitational 
action of general form
\begin{equation}
  S[g] = \int d^4x \, \sqrt{g} \left( \lambda - kR +
  \frac{1}{4} a R_{\mu \nu \rho \sigma} R^{\mu \nu \rho \sigma}
  \right) .
\label{azione}
\end{equation}

The constants $k$ and $\lambda$ are related -- in general 
as ``bare quantities'' -- to the Newton constant $G$ and to the 
cosmological constant $\Lambda$: $k$ corresponds to $1/8\pi G$
and $\lambda$ to $\Lambda/8\pi G$. The trajectories $x_i(t)$ of 
$m_1$ and $m_2$ are parallel with respect to the metric $g$; let
$R$ be their distance.

In the weak-field approximation, eq.\ (\ref{ciao}) reproduces
to lowest order the Newton potential and can be used 
to find its higher order quantum corrections \cite{muz}, or implemented 
on a Regge lattice to investigate the non-perturbative
behaviour of the potential at small distances \cite{h}. The 
addition to the gravitational action (\ref{azione}) of a term 
which represents a localized {\it external} Bose condensate 
\footnote{This means that the density of the condensate is not 
included into the functional integration variables.}
mimics a shielding effect which is absent from the classical
theory and which we take as our candidate model for the
observed shielding.

The feature of eq.\ (\ref{ciao}) which is of interest here 
is that if the two masses $m_1$ and $m_2$ are not pointlike,
the trajectories $x_1(t)$ and $x_2(t)$ must be those of their
centers of mass. (This also makes irrelevant the question --
actually ill-defined in general relativity --
whether they are pointlike or not.) Thus, when applying
eq.\ (\ref{ciao}) to the Earth and the sample, we only need
to consider the centers of mass of those
bodies. In this way we reproduce the observed behaviour for the 
shielding as well as for the regular interaction.
The ensuing apparent failure in the "local transmission" of the
gravitational interaction does not contrast with any known
property of gravity (compare \cite{m3,j}, and references about
the problem of the local energy density in General Relativity
and \cite{m3} about the non-localization of virtual gravitons.
One should also keep in mind that (\ref{ciao}) holds only
in the static case.)

Finally, if we describe the shielding effect as a slight
diminution of the effective value of the gravitational
acceleration $g$, and remember that the gravitational potential
energy $U=-\frac{G m_{Earth} }{r_{Earth} }=-g r_{Earth}$ 
is negative, it follows that the energy of a sample inside
the shielded zone is larger than its energy outside. This
means in turn that the sample must feel an expulsive force
at the border of the shielded region. Such a force has been
indeed observed \cite{p3}, although precise data are not
available yet. From the theoretical point of view it is
however not trivial to do any prevision about the intensity
of the force. In fact, the shielding process absorbs energy
from the experimental apparatus and thus any transient
stage is expected to be highly non-linear, especially for
heavy samples.

\end{document}